\documentclass{PoS}
\usepackage{sma-prop04sep}
\usepackage{amssymb}
\let\OLDthebibliography\thebibliography
\renewcommand\thebibliography[1]{
  \OLDthebibliography{#1}
  \setlength{\parskip}{1.4pt}
  \setlength{\itemsep}{1.6pt plus 0.3ex}
}

\title{Wind-driven Exclusion of Cosmic Rays in the Protoplanetary Disk Environment }

\ShortTitle{Wind-driven Exclusion of Cosmic Rays in the Protoplanetary Disk Environment }

\author{\speaker{L. Ilsedore Cleeves}, Edwin A. Bergin, Fred C. Adams%
         \\
        University of Michigan\\
        E-mail: \email{cleeves@umich.edu}}

\abstract{The recent (apparent) passage of the Voyager 1 spacecraft into interstellar space provides us with front-row seats to the complex interplay between the solar wind and the protective surrounding
bubble known as heliosphere. The heliosphere extends radially out to $\sim100$~AU from the sun,
and within this sphere of influence, the solar wind modulates the incoming flux of galactic cosmic
rays (CRs), especially those at low energies. Newly formed stars, which support both strong
magnetic fields and winds, are expected to produce analogous regions of CR exclusion, perhaps
at elevated levels. Such young stars are encircled by molecular gas-rich disks, and the net removal
of CRs from the circumstellar environment significantly reduces the expected CR ionization rate
in the disk gas, most likely by many orders-of-magnitude. The loss of ionization reduces disk
turbulence, and thereby affects both planet-formation and active chemical processes in the disk.
We present models of CR exclusion and explore the implications for turbulence and for predicted
chemical abundances. We also discuss means by which ALMA can be used to search for extrasolar
 heliosphere-analogs around young stars.
 }

\FullConference{Cosmic Rays and the InterStellar Medium\\
                 24-27 June 2014\\
                 Montpellier, France}

\begin{document}

\section{Introduction}
\begin{figure}
\begin{centering}
\includegraphics[width=0.89\textwidth]{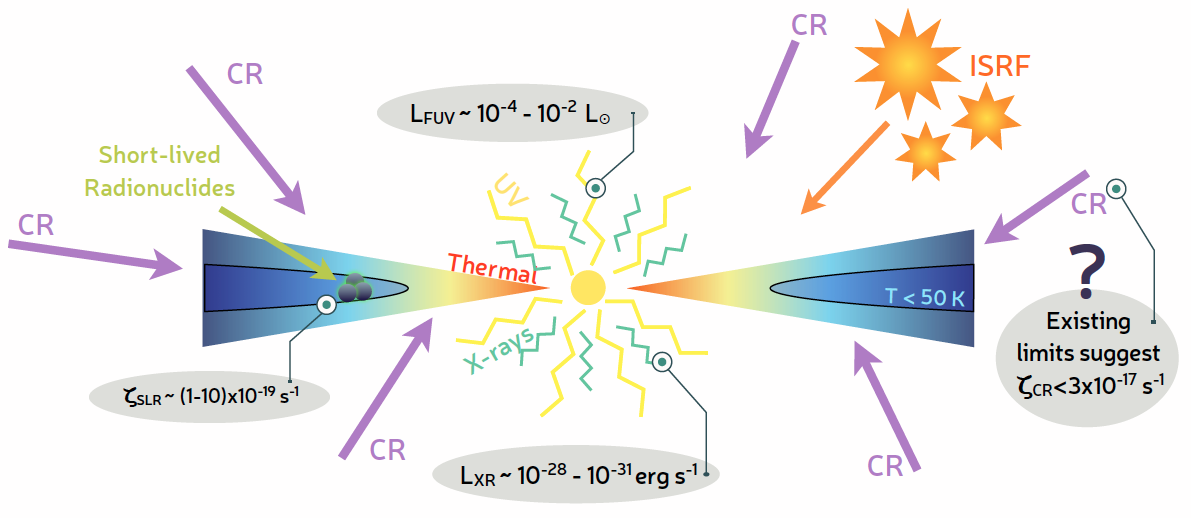} 
\caption{Sources of energetic ionization available to disks. The central star is both UV and X-ray bright.
The FUV photons both from the central star as well as neighboring stars (the interstellar radiation field or ISRF) ionize species with low ionization potentials, such as C, S, and Si. X-rays from the star directly ionize H$_2$ and He, but primarily in the upper layers that are directly exposed to the X-rays. Short-lived radionuclides are locked up within the dust and thus ionize the disk internally, but their influence gradually decays over time. Cosmic rays (CR) are thought to pervade the outer disk since they experience the least obstruction. Observational evidence for each of these sources is highlighted in grey. All sources except the cosmic ray ionization rate have been measured in disks, whereas cosmic rays have only upper limits.\label{fig:1}}
\end{centering}
\end{figure}

Ionization is fundamental to the chemical properties and physical structure of planet-forming
disks around young stars. The gas-rich disks around T~Tauri stars, analogues or the solar system
at an age of a few Myr, experience a diversity of ionization sources, both in magnitude and spatial
scope. Figure~\ref{fig:1} illustrates the main disk ionizing agents, along with characteristic values. In the
densest gas, the most important sources are stellar X-rays, short-lived radionuclide decay, and cosmic
rays. UV radiation from the central star or the potentially enhanced local UV environment (the
interstellar radiation field) may also contribute to the overall disk ionization but is fairly restricted
to an outer surface due to efficient attenuation of UV light by small dust particles. We have observational handles for many of these components: T~Tauri stars are measured to be X-ray luminous, typically $L_{\rm XR}\sim10^{28} - 10^{31}$~erg~s$^{-1}$ \cite{feigelson2002} and are often time-variable, brightening by factors of a few. We can look towards the meteoritic record for evidence of our own short-lived radionuclide abundance history, where $^{26}$Al was the dominant contributor towards SLR-derived ionization for the first few million years of the gas disk lifetime \cite{umebayashi1981,adams2010,dauphas2011}. However, we have relatively little information regarding cosmic ray ionization of protoplanetary disks. Existing observations indicate that the CR ionization rate is not high, with limits of $\zeta_{\rm CR} < 3 \times10^{-17}$~s$^{-1}$ determined from the non-detection of the dense gas ion-tracer H$_2$D$^+$ \cite{chapillon2011}.

There is some theoretical expectation that the cosmic ray ionization rate should be low. Cosmic
rays are easily deflected by large scale interstellar or local disk magnetic fields \cite{dolginov1994,cleeves2013a,padovani2013,fatuzzo2014}, or magnetized winds from the disk and/or the star itself \cite{gammie1996,chiang2007,turner2007,cleeves2013a}. In this contribution, we present models of elevated levels of stellar wind modulation of cosmic rays by young T~Tauri stars. The
region over which winds operate is analogous to our own heliosphere, i.e., a ``T~Tauriosphere.''
We place these prescriptions in context with other contributors to the dense gas ionization, namely
X-rays and short-lived radionuclide decay \cite{cleeves2013b}. We then make use of the sensitive link between the spatial structure of disk ionization and the chemistry of molecular ions to provide observational predictions. Deep submillimeter observations of molecular ions will help determine both the global ionization rate and potentially disentangle individual ionizing sources with spatial information.
In this context, young stars with gas rich disks may provide the best opportunity to detect and
probe the structure of extrasolar heliospheres, using the molecular content of the disk as a ``test
particle.'' Finally, we present applications of our model to the well-studied TW~Hya protoplanetary
disk system, where we have identified for the first time a substantially reduced CR rate in the
circumstellar environment of another (in this case, young) star.

\section{Cosmic Ray Modulation by Winds}
The modern day solar wind, with a relatively modest mass loss rate ($\dot{M}\sim10^{-14}$~M$_\odot$~yr$^{-1}$) and typically Gauss-strength surface magnetic fields, drives out low energy CRs ($E_{\rm CR} < 100$~MeV) from a $ \sim120$~AU region encompassing the solar system \cite{webber2013}. We can estimate the corresponding cosmic ray ionization rate for molecular hydrogen under modern day solar conditions by taking the measured cosmic ray proton spectra at heliocentric distances of $\sim1$~AU and the energy-dependent H$_2$ ionization cross section \cite{padovani2011}. Under the influence of relatively mild sun-like winds, the cosmic ray ionization rate drops by a substantial level, $\sim1-2$ orders of magnitude. In the environment of a young star, with kilo-Gauss stellar magnetic fields \cite{basri1992,johnskrull1999,yang2005} and measured mass loss rates typically 10\% of the mass accretion rate \cite{hartigan1995}, or $\dot{M} \lesssim 10^{-9}$~M$_\odot$~yr$^{-1}$ (i.e., five orders of magnitude higher than that of the sun) cosmic ray modulation should be even more severe. This realization motivated us to explore models under more extreme wind-modulation conditions, as would
be expected for the circumstellar environment of young, magnetically active stars. To estimate
the effects of enhanced young stellar winds, we start from the force-field approximation for the
solar cosmic ray spectrum \cite{gleeson1967}. This approximation is an empirically derived prescription that describes the overall shape of the interplanetary cosmic ray spectrum to high accuracy with a single free parameter, the modulation potential \cite{usoskin2005}. The modulation potential has units of energy and conceptually describes the ``barrier'' due to winds that impedes the propagation of cosmic rays. The barrier varies in strength substantially over the solar magnetic activity cycle, with values of
500 MeV to 1,500 MeV \cite{usoskin2005}. To approximately estimate the enhanced level of wind-driven exclusion by a T~Tauri star, we extrapolate these observations using the magnetic activity of the sun (traced by sunspot coverage) as our proxy. At the relatively high levels of magnetic activity typical
of Myr-old stars, the local cosmic ray ionization flux is strongly modulated such that the ionization
rate drops by an additional $\sim 1-2$ orders of magnitude below solar rates, resulting in net cosmic
ray ionization rates of just $\zeta_{\rm CR}\sim10^{-22} - 10^{-20}$~s$^{-1}$. Our simple extrapolated-modulation models agree with cosmic ray proton spectra from the literature for the Gyr-old sun \cite{svensmark2006,cohen2012}, which may imply that our estimates for the magnitude of wind-modulation by T~Tauri stars are too conservative, and that the actual reduction of cosmic ray flux by winds may be more extreme. The full range of wind-modulated cosmic ray spectra are shown in Figure 2, where the diversity of local interstellar spectra (LIS) of cosmic rays highlights the difficulty of determining the particle spectrum at low ($\lesssim100$~MeV) energies. The most important feature of the highly modulated T~Tauri CR spectra (TT~Min and TT~Max, shown in orange), is that CRs of all energies, even at $\sim$GeV, are deflected
in the presence of extreme winds.

\begin{figure}
\begin{centering}
\includegraphics[width=0.53\textwidth]{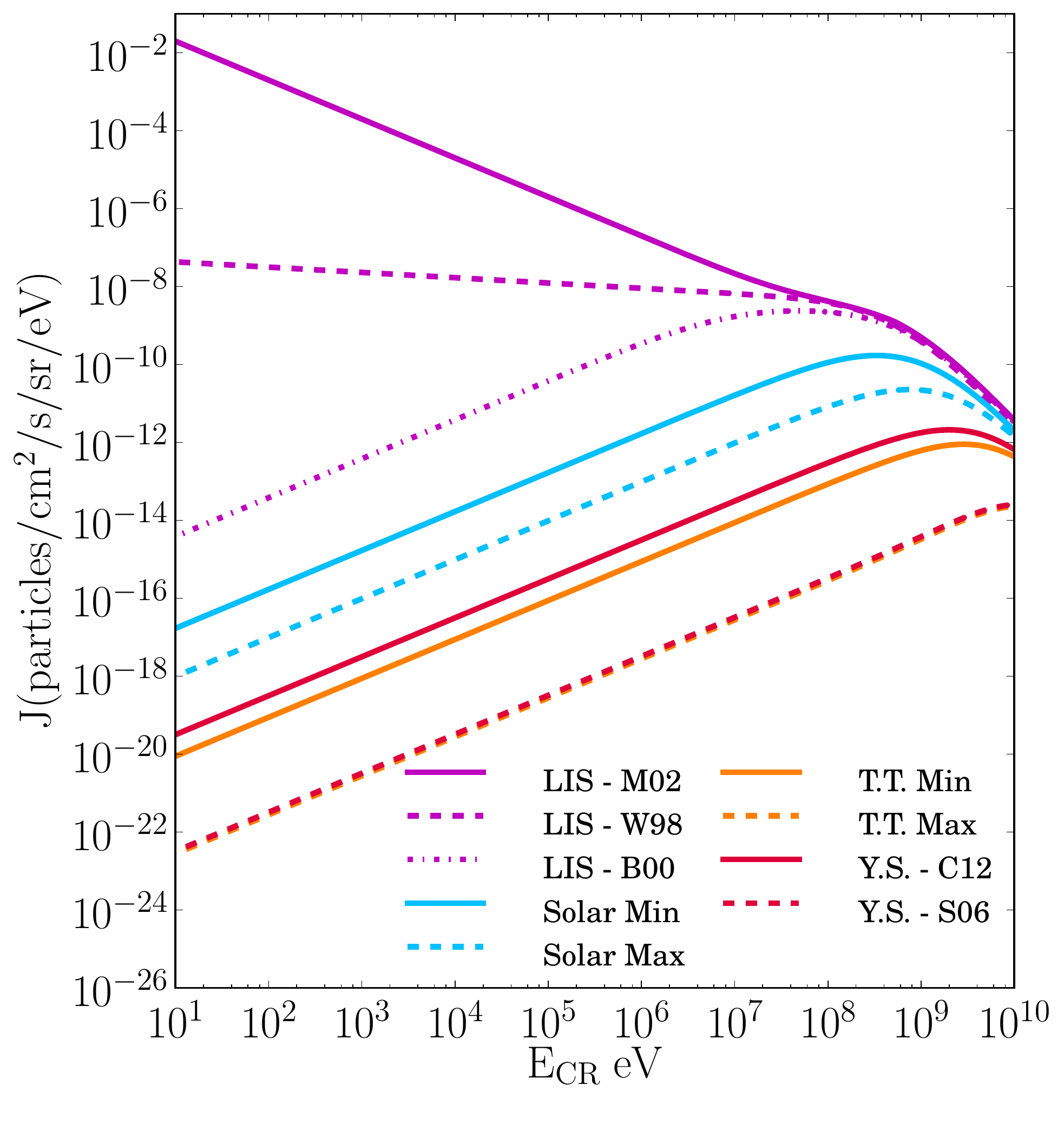} 
\caption{Cosmic ray proton spectra under the influence of stellar winds. The local interstellar spectra (LIS) are not well constrained below 1~MeV, as a direct result of the solar wind. The Voyager~1 points in red are the most recent direct measurements of the LIS \cite{webber2013}. The solar wind (blue) reduces the low energy end of the cosmic ray spectrum by over four orders of magnitude below 10 MeV. The orange lines show our extrapolated model sun results, estimated for T~Tauri stars. These results agree with the relatively late-stage Gyr-old young sun models in red, indicating that the T~Tauri spectra are possibly too high. {\em Reproduced from Cleeves, Adams, and Bergin 2013, ApJ, 772, 5.}  \label{fig:2}}
\end{centering}
\end{figure}

Cosmic ray exclusion by winds from young stars should be extreme. However, the region of
influence affected by magnetized winds is unknown. If stellar winds are the source of the cosmic ray
deflection, then the winds may be diverted by the massive gas disk or be confined by stellar
and/or disk magnetic fields, forming a doubly-lobed T~Tauriosphere. This geometry would still
permit cosmic ray ionization at the outer exposed rim of the disk. Alternatively, if large-scale,
magnetized disk winds (e.g., centrifugally launched or photoevaporative winds) are able to drive out
cosmic rays, they may create a larger shielded region. For example, many of the proplyds in Orion
show wide-angle bow shocks not associated with a jet. These bow shocks have been attributed to
a photoevaporative flow from the disk (greatly enhanced by UV irradiation from nearby massive
stars) colliding with the stellar winds emanating from the neighboring massive young stars. In the
case of proplyd 180-311, the bow shock wraps around the propyld at a projected radial distance of
700~AU \cite{garciaarredondo2001}, encompassing roughly $\sim200$ times the volume of our own heliosphere. If such flows are able to modulate the cosmic ray flux similarly to the solar wind, they could fully enclose the disk and shield the entire disk from low energy cosmic rays.

\section{Alternative Sources of Disk Ionization}
In the absence of cosmic rays, X-rays and short-lived radionuclides are alternative sources of
H2 and He ionization in the very dense ($n_{\rm H_2}\sim10^{10}$~cm$^{-3}$) gas. The ionization energy available in X-rays is extremely large; however, the geometry of the star/disk system is such that the $\lesssim 3$~keV X-rays experience the highest levels of disk gas and dust absorption and are thus unable to efficiently
penetrate the bulk gas. Thus the power in X-rays is mitigated by the large absorption column, which
is the greatest at the inner disk near the central star, providing a ``shield'' for the outer disk. X-rays
have the greatest influence in a surface layer where the gas surface density is below $\Sigma_g \lesssim 1$~g~cm$^{-2}$ \cite{bethell2011a}. During stellar X-ray flares, however, the overall spectrum tends to harden, with relatively more energy emitted at $> 3$~keV \cite{skinner1997}. The harder X-ray photons are less efficiently stopped by the disk and are able to ionize more of the bulk disk mass. The relative importance of these time-dependent effects will depend on the strength and frequency of the stellar flares, as well as ion-recombination timescales within the disk.

Short-lived radionuclides provide a baseline to the dense gas ionization rate in the absence
of cosmic rays; however, the typical rates range between $\zeta_{\rm SLR}\sim 10^{-18}$~s$^{-1}$ in the inner disk to $\zeta_{\rm SLR}\sim 10^{-19}$~s$^{-1}$ in the outer disk, taking into account the loss of the energetic decay products from the lower density regions of the disk \cite{cleeves2013a}. This rate, furthermore, decays over timescales shorter than the disk lifetime, with a half-life of approximately 1.2~Myr. After 10~Myr (without late-stage resupply of radioactive materials via supernovae or Wolf-Rayet stars \cite{ouellette2007,ouellette2010,adams2014}), the decay rate is just $\sim0.3\%$ of its initial value, or $\zeta_{\rm SLR} \sim (1-10)\times10^{-22}$ s$^{-1}$.

\section{The Chemistry of Disk Ionization}
In light of these effects, it is clear that the global picture of disk ionization is highly complex.
Winds and magnetic fields introduce orders of magnitude reduction in interstellar cosmic ray
rates. X-rays provide a floor to the ionization rate of about $\zeta_{\rm XR} \sim (1-10)\times10^{-21}$~s$^{-1}$ (varying with disk mass), but hard X-ray flares enable deeper X-ray propagation, which can increase the instantaneous X-ray ionization rate by orders of magnitude during the flare. Likewise, the inherent time-dependence (and supply history) of the abundance of short-lived radionuclides like $^{26}$Al also introduces at least an order of magnitude uncertainty in its contribution to the overall disk ionization budget \cite{finocchi1997,umebayashi2009,cleeves2013b}.

Fortunately, the chemistry of molecular ions is highly sensitive to the underlying ionization
environment. We explored a large grid of ionization models to calibrate the sensitivity of specific
molecules and particularly specific emission lines to variations in the ionization source and magnitude
\cite{cleeves2014par}. Examples of the dependency are shown in Figure~3. HCO$^{+}$ traces very high cosmic ray ionization rates, but is insensitive to low rates once the cosmic ray contribution falls below the X-ray contribution in the regions where HCO$^{+}$ is formed. N$_2$H$^{+}$ and N$_2$D$^{+}$ are particularly sensitive tracers of the deep ionization since these species are preferentially destroyed in warm gas, above 20~K, due to CO evaporation.

\begin{figure}
\begin{centering}
\includegraphics[width=0.72\textwidth]{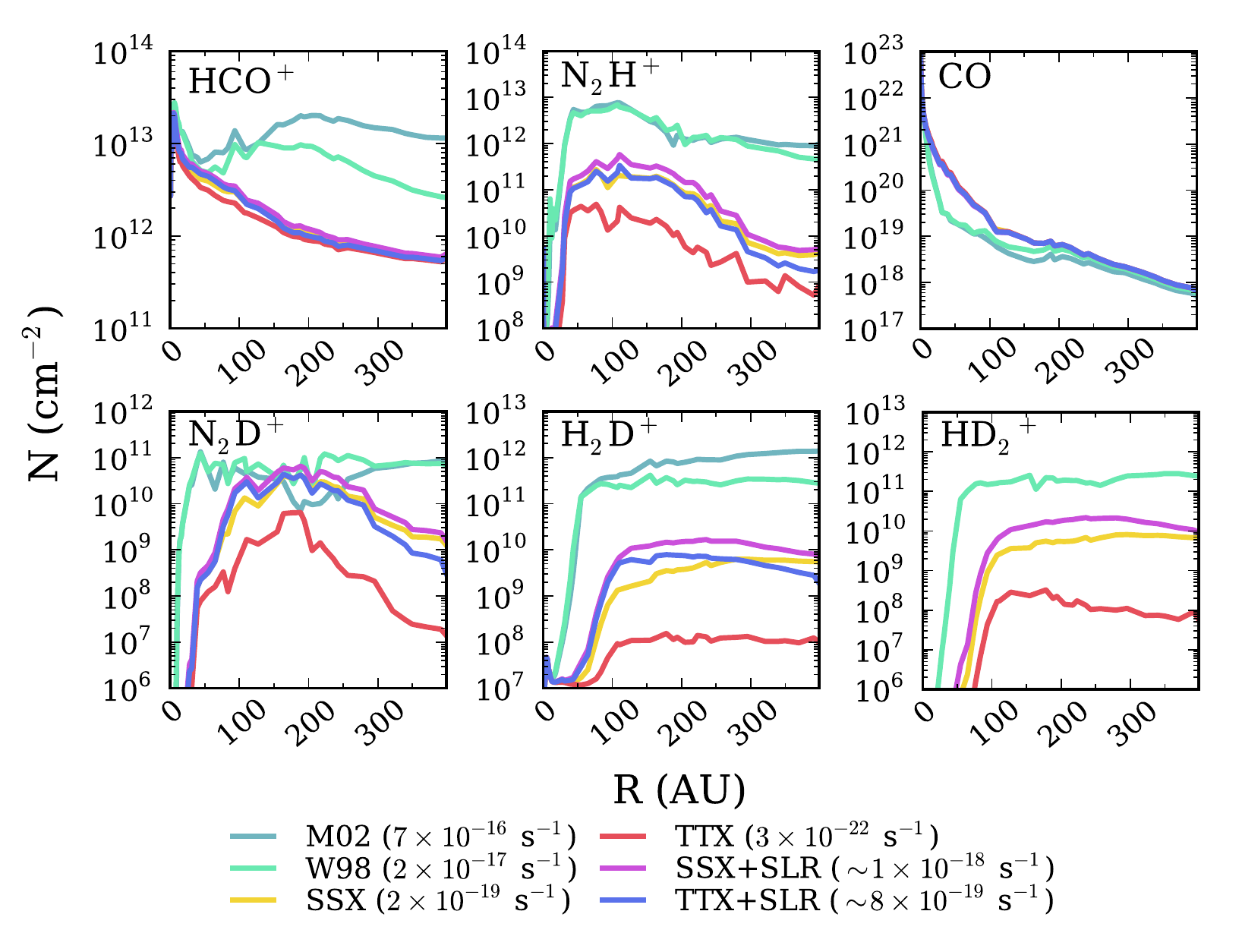} 
\caption{Variations in the column densities of important molecules for different cosmic ray ionization rates, with and without short-lived radionuclide ionization for the lowest CR ionization cases. These chemical signatures can be used in conjunction with observations to infer the ionization rate in T~Tauri systems. {\em Adapted from Cleeves, Bergin, and Adams 2014, ApJ, 794, 123.}  \label{fig:3}}
\end{centering}
\end{figure}

We applied these results to one of the most extensively studied protoplanetary disks, TW~Hya.
TW~Hya is relatively close ($d=55\pm9$~pc) \cite{webb1999,vanleeuwen2007} and face on with an inclination of only $i\sim7\pm1^\circ$ \cite{qi2004}. Starting from a well-calibrated model of the bulk gas disk, we explored twenty different ionization model combinations. We primarily varied the X-ray spectral hardness ratio (corresponding to small changes in X-ray luminosity, less than a factor of $\sim3$) and the incident cosmic ray flux using the wind models described above. For each model combination, we calculated the resulting disk chemical structure, and then simulated the emergent line emission for lines of HCO$^+$, H$^{13}$CO$^+$, and N$_2$H+, comparing seven independent observations of molecular ion lines in total. The data were systematically fit better with low cosmic ray ionization rates, with the best fit models falling below $\zeta_{\rm CR} \lesssim 10^{-19}$~s$^{-1}$. These results indicate that a significant degree of cosmic ray modulation is active in TW~Hya, suppressing the overall ionization rate by two orders of magnitude or perhaps more. The actual cosmic ray rate could actually be much lower, because the measured rate appears to reflect TW~Hya?s stellar X-ray ionization floor. These results put strong limits on the combined cosmic ray and short-lived radionuclide ionization rates in the TW~Hya protoplanetary disk. If winds are the primary driver of cosmic ray exclusion, these results may highlight the first detection of an external heliosphere, extending beyond 200~AU, i.e., beyond the molecular gas disk.

\section{Conclusions}
We have presented models that produce highly modulated cosmic ray ionization rates due to
the influence of stellar winds from a young, Sun-like (T~Tauri) star. We put the derived ionization
rates into context by comparing with those of X-rays and radionuclides. Variations in all sources
of ionization induce observational signatures in the molecular ion chemistry of disks, in particular, in the radial abundance profiles. Applying this knowledge, we have determined a substantially
reduced cosmic ray ionization rate in TW~Hya \cite{cleeves2014sub}. In this particular
system, the ionization appears to be primarily X-ray driven, implying that it may be difficult
to further isolate the short-lived radionuclide and cosmic ray components without observations of
the X-ray shielded gas in the inner disk, which will soon be possible with the full resolution and
sensitivity of the Atacama Large Millimeter Array. With future observations and theoretical development,
we will be able to measure the chemical structure of T~Tauri disks, infer their ionization
rates, and constrain the initial conditions for planet formation.

\end{document}